\documentstyle[epsf]{mn}

\newif\ifAMStwofonts

\def\gtorder{\mathrel{\raise.3ex\hbox{$>$}\mkern-14mu
             \lower0.6ex\hbox{$\sim$}}}
\def\ltorder{\mathrel{\raise.3ex\hbox{$<$}\mkern-14mu
             \lower0.6ex\hbox{$\sim$}}}

\title[SDSS J131339.98$+$515128.3]{SDSS J131339.98$+$515128.3: A new gravitationally lensed quasar selected based on near-infrared excess} 

\author[Ofek et al.]
       {Eran~O.~Ofek\thanks{e-mail: eran@astro.caltech.edu}$^{1}$, Masamune~Oguri$^{2}$, Neal~Jackson$^{3}$, Naohisa~Inada$^{4}$, and Issha~Kayo$^{5}$ \\
        $^{1}$ Division of Physics, Mathematics and Astronomy, California Institute of Technology, Pasadena, CA 91125 \\
        $^{2}$ Kavli Institute for Particle Astrophysics and Cosmology, Stanford University, Menlo Park, CA 94025. \\
        $^{3}$ The University of Manchester, Jodrell Bank Observatory, Macclesfield, Cheshire SK11 9DL, UK. \\
        $^{4}$ Cosmic Radiation Laboratory, RIKEN (The Institute of Physical and Chemical Research), 2-1 Hirosawa, Wako, Saitama 351-0198, Japan. \\
        $^{5}$ Department of Physics and Astrophysics, Nagoya University, Chikusa-ku, Nagoya 464-8602, Japan}

\date{Accepted ?
      Received ?
      in original form ?}

\begin{document}

\maketitle

\begin{abstract}

We report the discovery of a new gravitationally lensed quasar, SDSS
J131339.98$+$515128.3, at a redshift of $1.875$ with an image separation
 of $1\farcs24$. The lensing galaxy is clearly detected in visible-light
 follow-up observations. We also identify three absorption-line doublets in
 the spectra of the lensed quasar images, from which we measure the lens
 redshift to be $0.194$. Like several other known lenses, the lensed
 quasar images have different continuum slopes.  This difference is
 probably the result of reddening and microlensing in the lensing galaxy.
 The lensed quasar was selected by correlating Sloan Digital Sky Survey
 (SDSS) spectroscopic quasars with Two Micron All Sky Survey (2MASS)
 sources and choosing quasars that show near-infrared (IR) excess.
 The near-IR excess can originate, for example, from the contribution of
 the lensing galaxy at near-IR wavelengths. We show that the near-IR
 excess technique is indeed an efficient method to identify lensed
 systems from a large sample of quasars. 

\end{abstract}


\section{Introduction}
\label{Introduction}

Construction of large samples of gravitational lenses has been shown 
to be a unique tool for studying galaxy mass profiles
(e.g., Cohn et al. 2001; Rusin et al. 2002; 
Koopmans \& Treu 2003; Rusin \& Kochanek 2005; Treu et al. 2006),
galaxy evolution (e.g., Kochanek et al. 2000; Ofek et
al. 2003; Treu \& Koopmans 2004; Rusin \& Kochanek 2005), and cosmology
(e.g., Linder 2004; Oguri 2007). Moreover, individual lenses enable us
to study the lensed quasar engine (e.g., Kochanek et al. 2007) and the
dark matter content of galaxies through microlensing (e.g., Wambsganss
et al. 2000; Schechter \& Wambsganss 2002). 

The photometric and spectroscopic quasar samples of
the Sloan Digital Sky Survey (SDSS; York et al. 2000) were estimated to
contain $\sim1000$ lensed quasars. However, in practice, because of
seeing limitations, it is challenging to find most of these lenses. The
most successful search for lensed quasars in the SDSS (Oguri et
al. 2006) utilizes the morphological properties of SDSS quasars, as well
as colours of nearby objects, to  efficiently select lensed quasars with
image separation greater than $1''$ and flux ratio above
$\sim0.3$. Lensed quasars can also be discovered spectroscopically by
looking for a compound spectrum consisting of a galaxy and a quasar
(e.g., Johnston 2003).
Since these techniques locate only a fraction of lensed
quasars in the SDSS, it is of great importance to develop additional
approaches in order to find more lenses, especially lenses with image
separation below $\sim1''$. 

The efficiency of lens surveys can be enhanced by combining
multi-wavelength data. For instance, Kochanek et al. (1999) compared
optical and radio properties of wide-separation quasar pairs to argue
that they are binary quasars rather than gravitational lenses.
Jackson \& Browne (2007) proposed a lens search method that exploits the
small position differences between radio positions of lensed images and
optical positions of lensing galaxies. The method may
enhance the efficiency of lens searches by up to a factor of $\sim 10$.
Another approach, presented by Haarsma et al. (2005),
is to look for extended radio lobes which are
lensed by galaxies along the line of sight.
We note that the magnification bias for such multi-wavelength
surveys is more complicated than usual
(e.g., Borgeest et al. 1991; Ofek et al. 2002; Wyithe et al. 2003).

In this paper we present an approach for finding lensed quasars in
the SDSS data. This method relies on selecting, as candidate lens
systems, quasars with near-IR excess, which may be due to the
contribution of a lensing galaxy.
We note in this context that Gregg et al. (2002) suggested that red
quasar samples may contain a large number of lensed quasars due to
reddening by the lensing galaxy. We show that the quasar
SDSS~J$131339.98+515128.3$ selected by this method is a
new gravitationally lensed quasar, which is confirmed from imaging and 
spectroscopic follow-up observations.

The outline of this paper is as follows. We describe the near-IR excess 
selection method and the lensed quasar candidates in \S\ref{Method}. In
\S\ref{Obs} we present the observations of the new lens we have found,
and its spectroscopy and photometry are described in \S\ref{Phot}. We
model the new system in \S\ref{Model} and discuss the results in
\S\ref{Disc}. Finally the conclusions are given in \S\ref{Conc}.

\section{Selection of lensed quasar candidates based on the near-IR excess method}
\label{Method}

\begin{figure}
\centerline{\epsfxsize=8.5cm\epsfbox{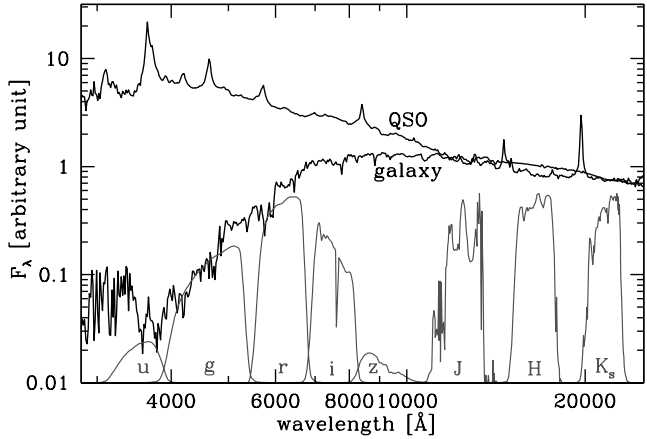}}
\caption{Illustration of the spectra of an elliptical galaxy (Mannucci
 et al. 2001) at a redshift $0.5$ and a quasar (Vanden Berk et al. 2001)
 at a redshift of $2.0$. The spectra are normalized such that the
 $i$-band magnitude difference is $i_{\rm gal}-i_{\rm QSO}=1$, which is
 typical for lensed quasar systems. The transmission curves of the SDSS
 $ugriz$ and the 2MASS $JHK_{s}$ filters are also shown (Moro \& Munari 2000).}
\label{Example_GalQSO_Spec}
\end{figure}

The selection process for gravitational lenses presented here is based
on the fact that most lenses are red galaxies whereas quasars are
much bluer. Therefore, even if the flux from the lens galaxy does not
dominate at visible-light frequencies, which is quite common among lens
systems discovered to date, it is expected that in some cases the
lensing galaxy has a non-negligible contribution to the integrated
brightness of the lens system at near-IR frequencies. This suggests that 
some of the lens systems may be characterized by a near-IR excess
that originates from the lensing galaxy. Figure~\ref{Example_GalQSO_Spec}
illustrates this idea by showing template spectra of quasars and
early-type galaxies redshifted to typical source and lens redshifts, respectively.
This plot clearly indicates that the lensing galaxy can modify
the near-IR to visible light colour of lensed quasar systems.
In this paper, we adopt the $g-H$ colour as a measure of the near-IR excess.

In order to exploit this idea to look for new lensed quasars in the
SDSS, we cross-correlated the SDSS Data Release 5 (DR5) spectroscopic
quasars\footnote{In this paper we do not use the DR5 quasar catalog of
Schneider et al. (2007), because this work began before
the public release of their catalog.} with the Two Micron All Sky Survey
(2MASS) point sources (Cutri et al. 2003). Specifically we looked for a
2MASS point source within $5''$ from each quasar position and extract
its IR magnitudes\footnote{We make this catalog available to the
community through the Vizier service:
http://vizier.u-strasbg.fr/cgi-bin/VizieR. The online catalog includes
also cross correlation with the FIRST radio catalog (Becker et
al. 1995).}. 
For our lens search, we have constructed a subsample of candidates by
selecting quasars with: redshift $3>z>1$; redshift confidence
z\_conf $>0.9$; $i$-band magnitude brighter than 19.0; and angular
distance between the SDSS and 2MASS sources smaller than 
$1\farcs5$. We excluded low-redshift quasars because their fluxes 
are sometimes affected to a large extent by contributions 
from host galaxies. We also avoided selecting high redshift quasars
since the lens galaxies of such quasars will be typically at high
redshifts, and therefore are too faint to be detected in the 2MASS
survey (see \S\ref{MethodNearIR}).
The smaller
angular distance cutoff (i.e., $1\farcs5$)
selects $95\%$ of all the
SDSS-2MASS sources found up to $5''$ apart. We use the relatively small
offset in order to remove well separated quasar-galaxy pairs. The
resulting number of SDSS-2MASS quasars is 3132.
\begin{figure}
\centerline{\epsfxsize=8.5cm\epsfbox{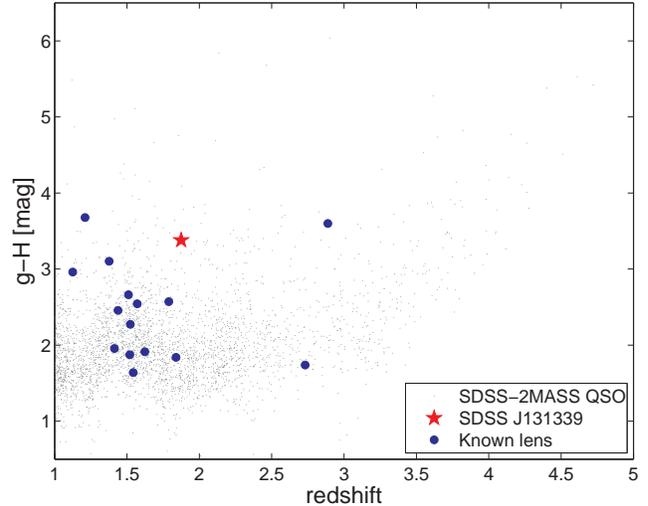}}
\caption{The optical-IR colour $g-H$ versus redshift for 3347 ($=3132+215$)
SDSS-2MASS quasars (dots) with SDSS measured redshifts above 1,
 redshift confidence $>90\%$, $i<19.0$~mag, and angular distance between the SDSS and
 2MASS sources smaller than $1\farcs5$. Also shown are the positions of known
lensed quasars (filled circles) and of SDSS~J$131339.98+515128.3$ (pentagram).}
\label{SDSS2MASS_gH_z}
\end{figure}
\begin{figure}
\centerline{\epsfxsize=8.5cm\epsfbox{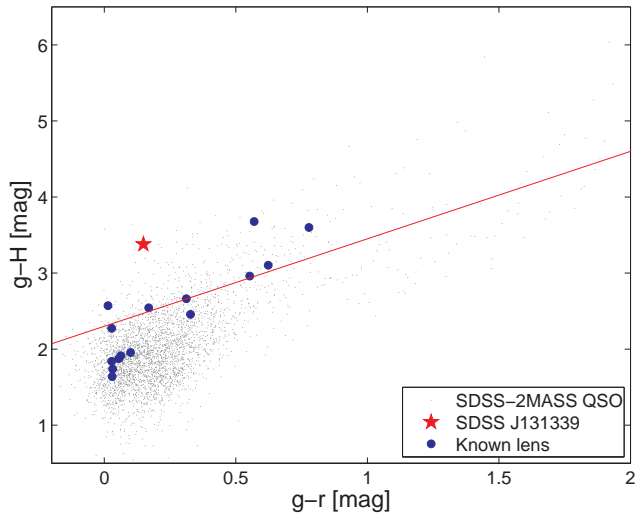}}
\caption{Same as Figure~\ref{SDSS2MASS_gH_z}, but for $g-H$ versus
 $g-r$. The line marks our selection cut in this colour space (see text).}
\label{SDSS2MASS_gH_gr}
\end{figure}
In Figures~\ref{SDSS2MASS_gH_z} and~\ref{SDSS2MASS_gH_gr}
we show the locations (dots) of the 3132 SDSS-2MASS quasars passing these
selection criteria, in addition to 215 
quasars found above redshift 3,
in the redshift versus $g-H$ colour and $g-r$
versus $g-H$ colour phase spaces, respectively.
Also shown are the positions of known lensed quasars with SDSS and
2MASS detections (filled circles), and the position
of the new lens reported in this paper (SDSS~J131339.98$+$515128.3;
pentagram). 
Figure~\ref{SDSS2MASS_gH_gr} indicates that the optical-IR colour $g-H$
is correlated with the optical colour $g-r$, reflecting the fact that the
quasar spectral energy distributions are characterized by a power-law.
The Figure suggests
that known lenses tend to be redder than the ``average'' quasar.

In order to test the feasibility of this method, from the 3132
quasars, we selected objects whose colours satisfy:
$g-H>1.15(g-r)+2.3$ (above the line in
Figure~\ref{SDSS2MASS_gH_gr}),
where, $g$ and $r$ are the $g$ and $r$-band 
magnitudes corrected for Galactic extinction (Schlegel et al. 1998;
Cardelli et al. 1989).
This colour cutoff is arbitrary  and does not rely on any
quantitative argument.
We choose this cut as an initial trial to study
the feasibility and efficiency of the near-IR-excess lensed quasar selection method.
Using the $g-H$ cut we selected
294 quasars, of which the following 7 are previously known lenses:
HS0810+2554 (Reimers et al. 2002);
SBS0909+523 (Kochanek et al. 1997; Oscoz et al. 1997);
SDSS~J1155+6346 (Pindor et al. 2004);
SDSS~J1206+4332 (Oguri et al. 2005);
SDSS~J1226$-$0006 (Inada et al. in preparation);
SDSS~J1335+0118 (Oguri et al. 2004);
SDSS~J1524+4409 (Oguri et al., in preparation).
Therefore the lens search efficiency is already $>2\%$ at
this point, before introducing any morphological constraints.

We make the lens survey more efficient,
and reduce the number of targets for follow up observations,
by applying a simple cut on the
morphology of the quasars, in the same spirit as Oguri et al. (2006).
Specifically, we set the condition that the SDSS stellar likelihood
of the candidates is
smaller than $0.1$ in both the $r$ and $i$-bands. The low stellar
likelihood of $<0.1$ means that the object is poorly fitted by the point
spread function (PSF).
We note that the completeness for including extended quasars
in the spectroscopic quasar sample begins to
decrease at $z=2.2$, and at $z\gtorder3$ most extended
quasars are not targeted (Oguri et al. 2006).
%
We finally compiled a list of 56 candidates
that survived the morphology cut (14 of them with $z>2.2$).
We note that all the 7 previously known lenses pass the colour and morphology
criteria.
Therefore, the lens selection efficiency improves to more than $10\%$.

Next, we look for new gravitationally
lensed quasars in this list of candidates.
We exclude seven of the radio-loud targets in
the list, because they were observed at $0\farcs22$ resolution 
in the CLASS radio lens survey
and confirmed not to be lenses (e.g., Myers et al. 2003).
We observed two targets at the Subaru 8.2\,m telescope
and 12 targets at the University of Hawaii 2.2\,m telescope, and
found that one of the targets, SDSS~J$131339.98+515128.3$, consists
of two optical point sources and an extended source, making it an excellent lens
candidate. SDSS~J$131339.98+515128.3$ has no detectable radio flux at
the $\sim1$~mJy limit of the FIRST 20-cm survey (Becker et al. 1995).


\section{Observations of SDSS~J131339.98$+$515128.3}
\label{Obs}

\begin{figure}
\centerline{\epsfxsize=8.5cm\epsfbox{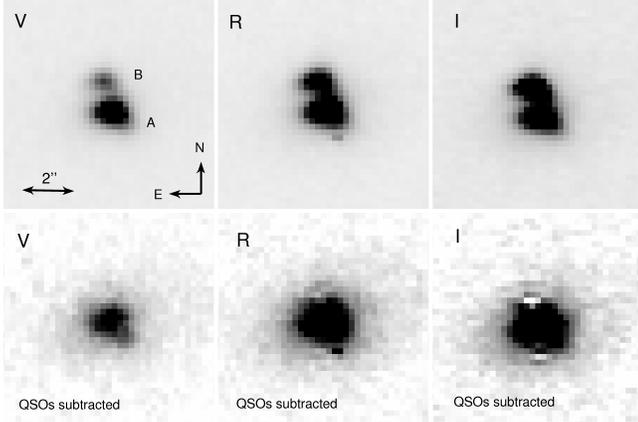}}
\caption{{\bf Upper panels:} $VRI$ frames (from left to right)
around SDSS~J$131339.98+515128.3$, obtained at
the University of Hawaii $2.2$~m telescope. The frames are presented
with the same gray-scale stretch, and the colour difference between the
two quasar images is apparent. Throughout this paper
the brighter quasar image (south) is denoted A and the fainter
(north) is denoted B.
{\bf Lower panels:} $VRI$ frames (from
left to right) of the lensed system after subtracting the two quasar
images. The lensing galaxy is clearly detected. The frames are displayed
with the same gray-scale stretch, which is about $1/6$ of the
gray-scale stretch in the upper panels. The white dots are residuals
due to the subtraction of the lensed quasar images.}
\label{SDSS1313_zoom}
\end{figure}

On UTC 2007 April 12.95, we observed our lens candidate
SDSS~J$131339.98+515128.3$ with
the Tektronix $2048\times2048$ CCD camera (Tek2k)
at the University of Hawaii $2.2$\,m  telescope, in $VRI$-bands. The 
exposure time was 300~s for each band.   
The frames with a pixel scale of $0\farcs 2195$~pix$^{-1}$,
were obtained under seeing of about $0\farcs8$, and
are shown in Figure~\ref{SDSS1313_zoom}.
The data were reduced using standard IRAF\footnote{The Image Reduction
and Analysis Facility (IRAF), is general purpose 
  software system for the reduction and analysis of astronomical
  data. IRAF was written by the programming group at
  the National Optical Astronomy Observatories (NOAO) in Tucson,
  Arizona. NOAO is operated by the Association of Universities for
  Research in Astronomy (AURA), Inc. under cooperative agreement 
with the National Science Foundation} tasks, and
photometric calibration was preformed using observations of the
Landolt (1992) standard star fields PG0918$+$029 and PG1633$+$099.
%
%

On UTC 2007 March 10.56 we obtained a 1400\,s spectrum of
SDSS~J$131339.98+515128.3$ using the Low Resolution Imaging
Spectrometer (LRIS; Oke et al. 1995) on the Keck-I 10\,m telescope.
In the red arm we used the 400/8500 grating, and on the blue side we
used the 400/3400 grism. The spectrum was obtained using the $0\farcs7$
slit, at a position angle of $16$\,deg, which goes through the two
lensed quasar images. For flux calibration we observed the
spectrophotometric standard star HZ~44. The spectra were reduced using
tools developed in the MATLAB environment (Ofek et al. 2006).

\section{Photometry and Spectra}
\label{Phot}

The positions and magnitudes of the lensed quasar images and lensing
galaxy are derived by fitting the system with two PSFs and a galaxy
profile using Galfit (Peng et al. 2002). We first model the galaxy by a
Sersic profile and found the best-fit Sersic index to be $n\approx 4$.
Therefore, we fix the index to $n=4$, a typical value for early type
galaxies, to compute the relative positions and magnitudes.
Tables~\ref{Tab-Pos} and \ref{Tab-Mag} summarize the fit results.
The separation of the  lensed quasar images  is $1\farcs24$, and the
flux ratios between the two quasar images are $3.63$, 
$1.67$ and $0.92$  in the $VRI$-bands, respectively.
\begin{table}
\begin{center}
\begin{tabular}{lcc}
\hline
Object & $\Delta\alpha$& $\Delta\delta$    \\
       & [arcsec.]     & [arcsec.]         \\
\hline
A & $ 0.0            $ & $ 0.0            $\\
B & $-0.325 \pm 0.002$ & $ 1.193 \pm 0.002$\\
G & $-0.064 \pm 0.009$ & $ 0.153 \pm 0.009$\\
\hline
\end{tabular}
\caption{The relative positions were calculated from the mean
of the $VRI$ image positions and the errors were estimated from the scatter.
The absolute position of image A is: 13:13:39.98  $+$51:52:28.37 (J2000.0).}
\label{Tab-Pos}
\end{center}
\end{table}
\begin{table}
\begin{tabular}{lcccccc}
\hline
Object  &  $V$  &  $R$  &  $I$  &  $b/a$  & PA    & $R_{e}$   \\
        & [mag] & [mag] & [mag] &         & [deg] & [arcsec.] \\
\hline
A & $18.26$ & $18.09$ & $17.45$ &        &         &        \\
B & $19.66$ & $18.65$ & $17.36$ &        &         &        \\
G & $18.74$ & $18.23$ & $17.49$ & $0.78$ & $-71$   & $0.75$ \\
\hline
\end{tabular}
\caption{The $VRI$ magnitudes of the lensed quasar images and the
lensing galaxy as obtained by two PSFs$+$galaxy fitting. The magnitudes are
given in the Vega system. We fitted a Sersic profile to the galaxy
using Galfit (Peng et al. 2002). The best fit Sersic index is $\sim4$.
$R_{e}$ is the effective radius of the galaxy.} 
\label{Tab-Mag}
\end{table}
We note that the lensing galaxy is also apparent in an $R$-band
image taken using the Suprime-cam on the Subaru telescope (Miyazaki et
al. 2002), which was retrieved from the Subaru telescope archive (Baba
et al. 2002). However we did not use the Subaru data for our analysis
because the PSF of the image is very unusual, which prevents reliable
photometry and astrometry of the system. 

\begin{figure}
\centerline{\epsfxsize=8.5cm\epsfbox{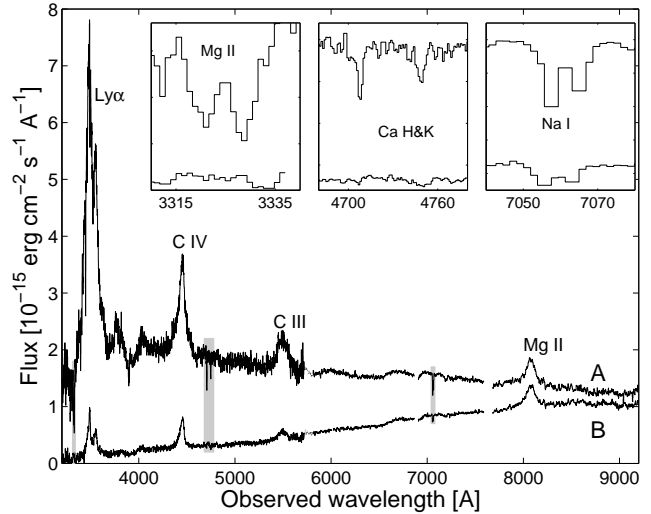}}
\caption{The spectra of SDSS~J$131339.98+515128.3$
images A (upper spectrum) and B (lower spectrum).
The three insets show absorption lines of
Mg~II ($\lambda\lambda 2798.1$-$2802.7$~\AA\AA),
Ca~II ($\lambda\lambda 3934.8$-$3969.6$~\AA\AA), and
Na~I ($\lambda\lambda 5889.9$-$5895.9$~\AA\AA),
in the spectra, presumably due to the lensing galaxy (see
Figure~\ref{SDSS1313_zoom}). The gray boxes mark the position of the
insets in the spectrum. The gaps in the spectra are due to removal of 
telluric absorption lines.}
\label{SDSS1313_SpecMain}
\end{figure}

\begin{figure}
\centerline{\epsfxsize=8.5cm\epsfbox{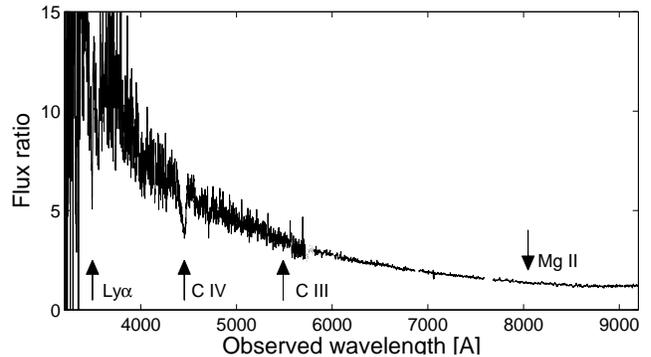}}
\caption{The ratio between the spectra of SDSS~J$131339.98+515128.3$
images A and B. The positions of the quasar broad emission lines
are marked by arrows.}
\label{SDSS1313_SpecRatio}
\end{figure}

The spectra of the two images of SDSS~J$131339.98+515128.3$
are displayed in Fig~\ref{SDSS1313_SpecMain},
and the flux ratio, as a function of wavelength,
between the spectra of images A and B is
presented in Fig.~\ref{SDSS1313_SpecRatio}.
The two quasars, marked by A and B, have identical redshifts with
similar (though not identical) 
broad emission lines, but markedly different continua.
We will discuss the possible
origins of this large colour difference in \S\ref{Disc}.
In the spectra we detect three absorption doublets of Mg~II,
Ca~II and Na~I (insets in Figure~\ref{SDSS1313_SpecMain})
at a redshift of $0.194\pm0.005$, which are probably induced by
the lensing galaxy. Additional support for this interpretation comes 
from the fact that the stellar absorption features of Ca H\&K are
stronger for image A, which is closer to the lensing galaxy.

We fitted the Galactic extinction (Schlegel et al. 1998) 
corrected colours of the galaxy with synthetic photometry
of galaxy spectral templates (Kinney et al. 1996) assuming the lens
redshift of $0.194$. The visible-light colours of the lensing galaxy
are best fit with a late-type galaxy.
However, these colours may be affected by the light of
image A, which is only $0\farcs17$ from the
position of the galaxy.
The late-type galaxy colours may contradict the
best-fit Sersic index of $\sim 4$.
To explore this issue, we tried fitting
the lens galaxy with an  $n=1$ Sersic profile, and found
that the fit to the imaging data is reasonable.
Therefore, further observations are needed for a more conclusive
classification of the lensing galaxy. 

\section{Lens modeling}
\label{Model}

We fitted, in the source plane, the image positions with an elliptical
softened power law mass model (Barkana 1998). In the fit we assumed a
core radius of 1 milliarcsecond (mas)
and used the following priors:
the lens position to lie within $27$~mas from the galaxy center (about $3$
times the errors in the galaxy position);
the lens eccentricity to be less than $0.35$;
the position angle (PA) to lie between $-40$ and $-100$ deg
(which is $\sim30$~deg from the PA of the galaxy light distribution);
and the power-law of the two-dimensional mass
distribution, $\alpha$, to be $1.0$ (corresponding to an isothermal sphere). 
The best fit parameters are: centre-of-mass position
$\Delta$R.A.$=-0\farcs048$, $\Delta$Dec.$=0\farcs175$ (relative to image A);
eccentricity $0.16$; PA$=-70$~deg; and the position of
the source is $\Delta$R.A.$=-0\farcs163$, $\Delta$Dec.$=0\farcs595$.
The time delay, estimated using the analytic formula for an isothermal
potential (Kassiola \& Kovner 1993), is around 30 days, assuming the third year
WMAP cosmology\footnote{$H_{0}=70.4$~km~s$^{-1}$~Mpc$^{-1}$; $\Omega_{m}=0.27$;
$\Omega_{\Lambda}=0.73$.} (Spergel et al. 2007).

Given these parameters, the derived magnifications are
$-0.4$ and $2.5$ for images A and B, respectively.
These predicted macro-model magnifications are inconsistent with the
observed flux ratio (i.e., the nearest image to the lens is the
brightest), and this discrepancy is discussed in the next section.

\section{Discussion}
\label{Disc}

Here we discuss the efficiency of our near-IR excess method
(\S\ref{EffNearIR}), its selection effects (\S\ref{MethodNearIR}),
as well as the nature of the different colours of the lensed images of
SDSS~J$131339.98+515128.3$ (\S\ref{DiscColor}).

\subsection{Efficiency of the near-IR excess method}
\label{EffNearIR}

Our preliminary selection criteria in the $g-r$ versus $g-H$ colour space
selected 294 candidates, of which at least 8 are gravitational lenses. 
The efficiency $\ge8/294\sim 3\%$ is apparently higher than lensing rates in
quasar samples. For comparison, the lensing efficiency of the CLASS lens survey
was $\sim0.14\%$ (Browne et al. 2003), and about $1\%$ in
the Hubble Space Telescope snapshot survey (e.g., Maoz et al. 1993).
Therefore we conclude that our near-IR
excess cut indeed enhances the fraction of lenses in the sample.
After the introduction of the morphology selection criterion,
the efficiency boosts to $\ge8/56\sim 14\%$. However,
such a morphology cut may drop very small separation lensed quasars that
are barely resolved in the SDSS.

We expect the efficiency of this method to increase for intrinsically
fainter quasars in which the lensing galaxy has a higher probability of
dominating the system light.
%
We comment that a significant fraction
of false positive candidates
in our near-IR  excess sample are close
pairs of quasars and galaxies which are inevitably selected given the
philosophy of the method.

\subsection{Selection effects of the near-IR excess method}
\label{MethodNearIR}

The near-IR excess selection method is biased toward systems in which
the lensing galaxy is apparently bright, and therefore, on average at
lower redshift than the typical redshift of lensing galaxies.
Moreover, these systems
will have larger than average image separations.
This is a consequence of the fact that
a brighter, and hence more massive, lensing galaxy implies
a larger image separation.
The lower lens redshift also
results in larger image separation.
%
Therefore, caution is needed when using lens systems found by such a
method in analysis which requires lensed quasar samples with
non-biased lens redshift distribution (e.g., Kochanek 1992; Ofek et
al. 2003; Capelo \& Natarajan 2007).

In addition, this method will probably be biased toward detecting doubly
imaged lensed quasars in which the average magnification is smaller than
in quadruply lensed systems, and therefore, the lensing galaxy has
a higher probability to dominate the total light in the near-IR.
We note that in a red-band flux-limited sample, like
the one we use here (i.e., $i<19$~mag), the incidence of lenses
will be increased due to the contribution of the lens
galaxy to the total system light,
relative to a sample which is not flux limited.

The images of some lensed quasars show remarkable rest-frame UV
continuum differences (e.g., HE1104$-$1805 Wisotzki et al. 1993) that
are variable with time (e.g., Ofek \& Maoz 2003; Poindexter et al. 2007).
Such continuum differences can make them appear redder.
Although this type of lensed quasars is relatively rare, it is
possible that the new quasar presented in this paper belongs to this group.
Furthermore, it is well known that narrow and broad absorption line QSOs
are redder than typical quasars (e.g., Brotherton et al. 2001), and this
is especially true for radio selected quasars (e.g., Menou et al. 2001).
Therefore, the near-IR excess method may be
efficient for looking for lensed broad absorption line quasars, which
may be useful for probing the structure of quasar outflows using
microlensing (Chelouche 2005).

Gregg et al. (2002) argued that the population of very red quasars 
contains a large fraction of gravitational lenses because of reddening
by the lensing galaxies.
Although we compare optical-IR colours ($g-H$) of quasars with the optical
($g-r$) colours in order to distinguish between red/reddened quasars and
lensed quasar systems, it is still possible that this method is biased
toward selecting red lensed quasars. This is also suggested by the fact
that one of the lensed components of our new lens has a red colour.
Note, however, that highly extincted lenses are
relatively rare (e.g., Falco et al. 1999).

Finally, we note that a basic limitation of this approach is that it can
detect the lensing galaxy only if it is brighter than the limiting
magnitude of the near-IR survey that is being used. In particular it
should be emphasized that the current use of this method is limited by
the shallow detection limit of the 2MASS survey used here.
In the 2MASS survey $5L_{*}$ galaxies can be detected up to a redshift
of $\sim0.2$.
Thus, lensed systems with higher-redshift lensing galaxies
cannot be efficiently detected using this method. However,
future surveys
(e.g., UKIDSS, Lawrence et al. 2006; WISE, Mainzer et al. 2005)
can be used to
select lensing galaxies to redshifts up to $\sim 1.0$, making the
near-IR excess method much more useful.  
%
%

\subsection{The origin of the different colours of the lensed images}
\label{DiscColor}

The lensed images of SDSS~J$131339.98+515128.3$ are distinctly
different in their spectral continuum slopes, but the shapes and
equivalent widths of their broad emission lines are similar (though not
identical). 
Such colour differences are not common among lensed quasars,
although there are several known examples (e.g., HE1104$-$1805).
Two obvious possibilities are that the different colours
are caused by microlensing, or/and by
extinction and reddening in the
lensing galaxy (e.g., Falco et al. 1999; El{\'{\i}}asd{\'o}ttir et
al. 2006).

The equivalent widths of the Na~I absorption lines in the spectra
of images A and B (Figure~\ref{SDSS1313_SpecMain}), are
$\sim2.5$~\AA~and $\sim1.5$~\AA, respectively. 
The known correlation between Na~I absorption lines
and extinction (e.g., Turatto et al. 2003)
suggests that the lensed quasar images
are extincted
with $E_{B-V}$ in the range of $0.1$ to $1$~mag.

To estimate the relative extinction between
the two quasar images, we divided the continuum-subtracted
spectrum of image A by that of image B
(i.e., we used only the broad line flux),
and fit it with extinction curves (Cardelli et al. 1989).
%
%
The best fit extinction we have found is $E_{B-V}=0.5$~mag
(assuming $R_{V}=3.08$).
Next, we divided the spectrum of image A by
the de-reddened spectrum of image B (i.e., $E_{B-V}=0.5$~mag, $R_{V}=3.08$).
The result is roughly consistent with a power-law,
$\sim\lambda^{-1\pm0.1}$.
This indicates that extinction alone probably cannot
explain the entire difference between the spectra.
We note that if we use the entire spectra (continuum and lines)
to fit for a relative extinction curve between
the two lensed images, a reasonable fit is found
with $E_{B-V}\approx0.7$~mag.
However, this does not constitute a perfect fit,
and the residuals between the spectra
are not monotonic with wavelength, which is difficult to explain.

Contrary to extinction,
we expect microlensing 
to alter the spectral continuum of a
quasar image, but keep the broad emission
lines unchanged.
The reason for this is that
the approximate length scale over which microlensing effects are
expected to average out, is several times the microlensing Einstein
radius. The expected Einstein radius for 1~M$_{\odot}$
microlensing in this system is about $0.03$~pc (in the source plane),
which is of the same order of magnitude as the expected size
of the accretion disks of bright quasars.

\begin{figure}
\centerline{\epsfxsize=8.5cm\epsfbox{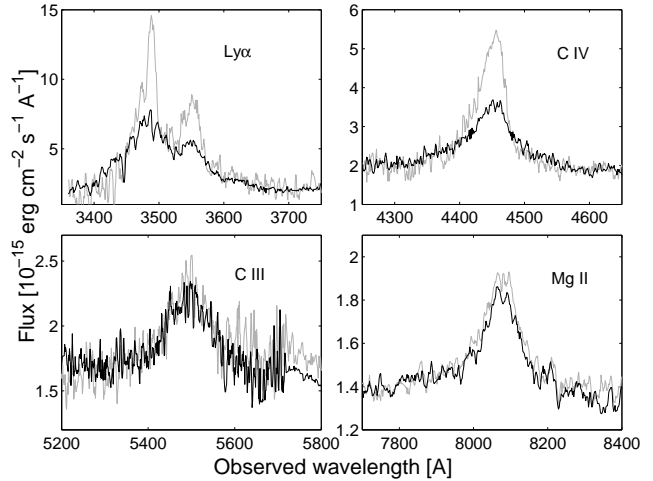}}
\caption{Each panel shows one of the major emission lines in the
spectrum: Ly$\alpha$, C~IV, C~III] and Mg~II, at the observed
wavelengths.
In each case,
the black-line shows the spectrum of image A,
and the gray line represents the observed,
extinction corrected ($E_{B-V}=0.5$, $R_{V}=3.08$),
spectrum of image B, multiplied
by 1.5, 1.0, 0.93 and 0.64,
for the Ly$\alpha$, C~IV, C~III], and Mg~II, respectively.
It can be seen that this scaling gives a good overlap for Mg~II and 
C~III] lines, and for the broad components of the other lines, but that there 
are narrow components in Ly$\alpha$ and C~IV lines,
which do not overlap well,
being considerably greater in image B.}
\label{SDSS1313_EmissionLinesComparison}
\end{figure}

Interestingly, as shown in
Fig~\ref{SDSS1313_SpecRatio},
the ratio between the lensed images' broad emission lines
is not identical to the ratio of the continua around these lines.
This can be investigated
further by a detailed comparison of the line profiles,
shown in Fig.~\ref{SDSS1313_EmissionLinesComparison}.
Here we have made the assumption that the
relative reddening between
images A and B is $E_{B-V}=0.5$~mag.
Then, for each broad emission line,
we have multiplied the de-reddened spectrum of image B
by a wavelength dependent factor,
%
%
in order to match the continua of images
A and B around these lines.
As shown in Fig.~\ref{SDSS1313_EmissionLinesComparison},
we find remarkably good agreement between
the lines of images A (black) and B (gray),
in the case of the C~III] and Mg~II lines.
In the other two lines (Ly$\alpha$ and C IV) the
broad wings of the line also agree, but there is clearly a narrower
component present within the broad emission lines of
image B that is absent in image A.

There are two possible explanations for this result. The first is that
the broader component of the emission lines
originates in a region small enough to be
microlensed, together with the optical continuum, in the sense that both
continuum and broad component of the emission lines in image
B are suppressed. 
We note that microlensing of a quasar's broad emission lines
was probably detected in SDSS~1004$+$4112
(Richards et al. 2004).
The second possibility is that reddening is mainly responsible for this
difference.
In this case we would
require that the continuum and broad line components
are affected by a different degree of reddening
than the narrow component of the emission lines.
In order for extinction in the lensing galaxy to produce a differential
effect of this type, it should be
patchy on a length scale of the broad-line region size (i.e., $\sim1$~pc). 
%
%
%

The most simple explanation for the spectral differences
between the lensed quasar images, is that while they
are affected by extinction, 
microlensing plays a significant role.
The spectral differences, after applying a
relative extinction of $E_{B-V}=0.5$~mag between the images,
suggests that the quasar region
responsible for the short-wavelength
($\lambda_{rest}\sim1400$~\AA)~continuum emission
is more magnified than the region emitting
the longer-wavelength 
($\lambda_{rest}\sim2800$~\AA)~radiation.
Indeed, this is consistent with what is expected from
microlensing of a quasar accretion disk.
Moreover, 
%
%
%
the negative parity image (image A in our case) has a higher probability
of being magnified by microlensing relative to the positive parity
image (image B; Schechter \& Wambsganss 2002; Mortonson et al. 2005).
Indeed, contrary
to the macro-model (\S\ref{Model}) prediction, image A is magnified
relative to image B. 
Probably the best way to verify that
microlensing is responsible for some of the
spectral differences between the lensed images, is to look
for temporal colour changes between the two lensed images,
similar to the colour variability observed
in HE1104$-$1805 (Poindexter et al. 2007).
%
%

\section{Conclusion}
\label{Conc}

Given the promising future of whole sky surveys (e.g., SkyMapper, Keller
et al. 2007; PanSTTAR, Kaiser et al. 2005; LSST, Tyson 2005), a variety
of methods to search for gravitationally lensed quasars are needed.
One possibility, suggested by Kochanek et al. (2006), is to look for
variable sources that are extended, presumably due to lensed quasar
images and/or lensing galaxy. However, this method, like the SDSS quasar
lens search (Oguri et al. 2006), is limited by the seeing of ground based
telescopes. 

We have proposed that some classes of lensed quasars can be identified
efficiently by looking for single-epoch near-IR excess of known
(spectroscopic or photometric) quasars. The near-IR method is based on
the fact that at least some lensed quasar systems will be redder due to
the presence of a bright lensing galaxy (e.g., Fig~\ref{Example_GalQSO_Spec}).  
The use of the near-IR excess method is demonstrated by our discovery of
the new lensed quasar SDSS~J$131339.98+515128.3$. Follow up observations
have shown it to be a doubly-imaged lensed quasar, at $z=1.875$, with an
image separation of $1\farcs24$. We also identified the lensing galaxy
at a redshift of $0.194$. The continua of the lensed quasar images are
markedly different, presumably due to extinction and microlensing effects.
The discovery of the new lens as well as the presence of many known
lensed quasars in the near-IR excess lens candidates sample suggests
that the near-IR excess is indeed an efficient way to select lensed
quasar systems from large samples of quasars.  

\section*{ACKNOWLEDGMENTS}
EOO, MO, and NJ would like to thank the Kavli Institute for
Theoretical Physics and the organizers of the KITP workshop
``Applications of Gravitational Lensing: Unique insights into Galaxy
Formation and Evolution'' for their hospitality; this work was initiated
at this KITP workshop.
It is pleasure to thank the referee, Chris Kochanek, for useful
comments on the manuscript.
This research was supported in part by the
National Science Foundation under Grant No. PHY05-51164, and also by
Department of Energy contract DE-AC02-76SF00515 and by the European
Community's Sixth Framework Marie Curie Research Training Network
Programme contract no. MRTN-CT-2004-505183 ``ANGLES''.  Use of the UH 2.2-m
telescope for the observations is supported by NAOJ.
Based in part on data collected at Subaru telescope and obtained
from the SMOKA archive, which is operated by
the Astronomy Data Center, National Astronomical Observatory of Japan. 
NI acknowledges support from the Special Postdoctoral 
Researcher Program of RIKEN.

\end{document}